# Design of achromatic surface microstructure for near-eye display with diffractive waveguide

Jiasheng Xiao[1,†], Juan Liu[1,†], Jian Han[2,*], and Yongtian Wang[1]

Dispersion problem has always constrained the development of see-through near-eye displays with diffractive waveguide. Here, we propose a design method and optical systems for achromatic surface microstructure composed with a triple-carved-sub-grating. The triple-carved-sub-grating with specified period is designed based on rigorous coupled wave theory, corresponding to Red, Green and Blue (RGB) wavelengths, respectively, with diffraction efficiencies of $\pm 1^{st}$ orders reach 47.5% under TE polarization. The surface microstructure with period of 18.9 $\mu m$ is realized by integrating the three sub-gratings together, and it is verified numerically that the diffractive angle of RGB wavelengths is 35° and the crosstalk is less than 4.5% under the normal incidence of TE polarization, which imply the surface microstructure diffracts the certain RGB wavelengths achromatically. Via proper duplication it could be utilized as the combining optics of diffractive-waveguide near-eye display and head-up display because of its light and compact features.

**Keywords:** dispersion; diffraction; microstructure; near-eye display

[1]China Beijing Engineering Research Center for Mixed Reality and Advanced Display, School of Optoelectronics, Beijing Institute of Technology, Beijing, 100081, China

[2]Key Laboratory of Information System and Technology, Beijing Institute of Control and Electronic Technology, Beijing, 100038, China

*These authors contributed equally to this work.

†Correspondence: Juan Liu, E-mail: juanliu@bit.edu.cn

## INTRODUCTION

See-through near-eye display (NED) is becoming a hot topic because of its various applications in both commercial and defense market[1-2]. By series of technical ways, researchers intend to make the NED as small and lightweight as the ordinary sunglasses[3-6]. Among all those technologies, using a transparent diffractive waveguide or diffractive optical element (DOE) as the optical combiner can dramatically reduce the weight and size of NED[7-10]. However, there still exist some problems which need to be solved, in aspect of imaging, the most important one is dispersion problem resulting from DOE when the NED displays a full-color image.

Here we propose a design method to eliminate the dispersion and optical waveguide system with a surface microstructure based on rigorous coupled wave theory, which diffracts RGB light achromatically. The surface microstructure could be served as the input/output couplers in diffractive near-eye displays to solve the problem of full-color display with only one stack of planar optical waveguide, which can dramatically reduce the weight and size of NEDs.

## MATERIALS AND METHODS

### General scheme

A sketch of the achromatic waveguide display system is shown in Figure 1. The RGB lights emitted from the micro display are collimated and coupled into the waveguide by the in-coupler, and then propagate along the waveguide in the condition of total internal reflection (TIR). When the lights hit the out-coupler, they are diffracted out of the waveguide and finally enter the human eye. The key part of this system is the achromatic in-coupler and out-coupler of waveguide, and we describe the basic design of the waveguide as follows.

The waveguide contains two microstructures carved on the top surface, and each of them is a triple overlapping etching grating, coupling light in or out of the waveguide without chromatic aberration. The design of the microstructure should be constrained by the following terms. (1) The diffraction angle should be greater than the TIR critical angle of the waveguide. (2)The structure of in or out coupler is same. (3) The couplers and waveguide use same material. (4) Three discrete wavelengths representing red, green, and blue light, respectively, are diffracted by the microstructure with the same angle. These constrains guarantee the achromatic microstructure can be applied to waveguide display and easily fabricated. Here, we choose grating as the basic component of the microstructure due to it can be designed to diffract light by a desired angle for one particular wavelength. If we use three sub-gratings to diffract three different wavelengths in same angle as well as satisfy the TIR angle of the waveguide, achromatic coupling will be realized. With appropriate design to overlap these gratings together, the coupler of the waveguide can be developed.

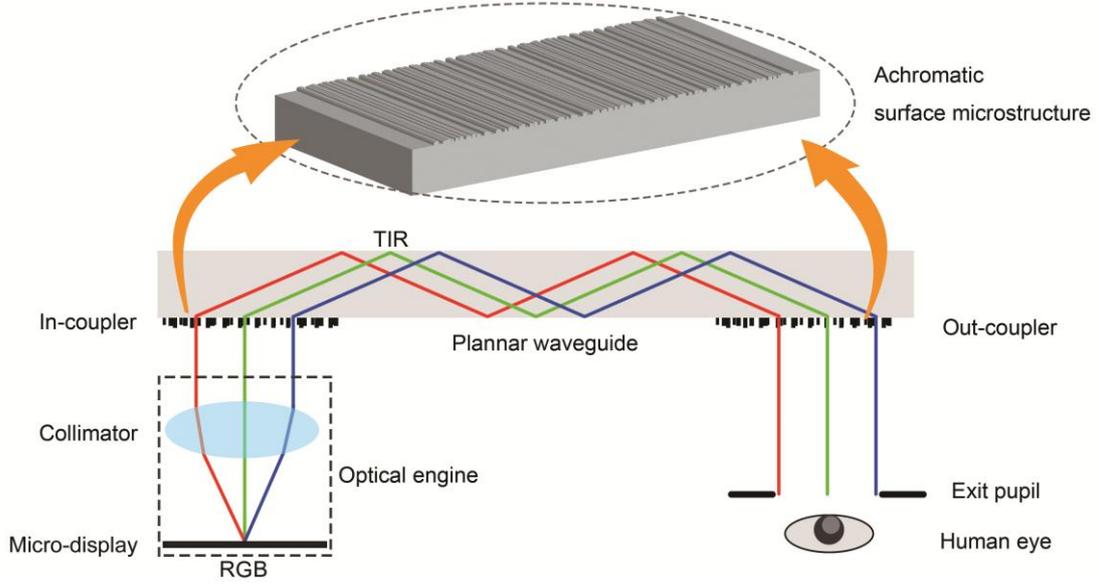

**Figure 1**. Scheme of diffractive near-eye display with achromatic surface microstructure. The red, green and blue lines refer to the optical paths of RGB wavelengths, respectively. Achromatic surface microstructure serves as the in-coupler and out-coupler, and it diffracts RGB wavelengths to the same diffractive angle

### Design theory and method

Rectangular gratings we choose as the basic component are easy to process for NEDs application, and they can be designed based on rigorous coupled wave theory[20] and grating equation. The transmission rectangular gratings for RGB wavelengths are described in Figure 2a, with period $\Lambda$, depth $d$, the reflective index of ridge $n_r$ and groove $n_g$, and duty cycle $f$. The incident light propagates in XZ plane with wavelength $\lambda$, the index of incident region is $n_g$ and the index of diffraction region is $n_r$. And grating equation is written as:

$$n_g \sin\theta - n_r \sin\theta_d = \frac{m_i \lambda_i}{\Lambda_i} \quad \text{for } i=R,G,B \quad (1)$$

Where $\theta$ and $\theta_d$ refer to the incident and diffractive angle, respectively, $m$ is the diffraction order, The diffraction efficiencies of the $i^{\text{th}}$ order reflected wave and transmitted wave under TE polarization and TM polarization are defined as:

$$\eta_{TERi} = R_i R_i^* \, \text{Re}\left(\frac{k_{I,zi}}{n_I k_0 \cos\theta}\right) \quad (2)$$

$$\eta_{TMRi} = R_i R_i^* \, \text{Re}\left(\frac{k_{I,zi}}{n_I k_0 \cos\theta}\right) \quad (3)$$

$$\eta_{TETi} = T_i T_i^* \, \text{Re}\left(\frac{k_{II,zi}}{n_I k_0 \cos\theta}\right) \quad (4)$$

$$\eta_{TMTi} = T_i T_i^* \, \text{Re}\left(\frac{n_I k_{II,zi}}{n_{II}^2 k_0 \cos\theta}\right) \quad (5)$$

Here $R_i$ is the normalized electric-field amplitude of the $i^{th}$ reflected wave, and $T_i$ is the normalized electric-field amplitude of the transmitted wave. Both of them are in relationship with period $\Lambda$, depth $d$ and duty cycle $f$. $k_{I,zi}$ and $k_{II,zi}$ are reflected and transmitted wave vectors along Z direction, respectively, which are defined as:

$$k_{l,zi} = \begin{cases} \sqrt{(k_0 n_l)^2 - k_{xi}^2} & k_{xi}^2 < (k_0 n_l)^2 \\ -j\sqrt{k_{xi}^2 - (k_0 n_l)^2} & k_{xi}^2 > (k_0 n_l)^2 \end{cases}, \quad l = I, II \quad (6)$$

where $k_{xi} = k_0 \left[ n_I \sin\theta - i(\lambda_0/\Lambda) \right]$.

According to the theoretical equations and design flow above, we use MATLAB to build an optimization model and find the best solution by varying the ridge height $d$ and duty cycle $f$ of sub-gratings.

**RESULTS AND DISCUSSION**

In order to solve the dispersion problem in NEDs, we designed three sub-gratings based on rigorous coupled wave theory. After all the three sub-gratings are calculated separately, surface microstructure can be obtained by superimposing them together.

Finite-difference time-domain (FDTD) simulations were performed to analyze its diffraction characters, as shown in Figure 4. The real part of near-field distribution is shown in Figure 5a, 5c, and 5e. Since the microstructure is designed for NEDs, by implementing Fourier transform of the near-field distribution, we can obtain its far-field distribution, which is shown in Figure 5b, 5d, and 5f. In figure 5b, 5d, and 5f, the far-field distribution suggested that the far-field intensity of $p$ polarization is almost zero and the diffraction angle of RGB bands is still 35° as desired. While for the red and blue bands, there exist weak diffraction crosstalk at 11° and 72.9° (see the black solid circle in Figure 5b and 5f) for $s$ polarization, but the main intensity is still concentrated at the diffraction angle of 35°.

The simulation results showed that achromatic diffraction is realized by this microstructure. Comparing with the previous methods[13-14] focusing on full-color near-eye displays, the surface microstructure proposed is one stack composed with one triple-carved-sub-grating, and its periodic structure is completely different from others, which are etched on the same surface of planar waveguide by integrating. This offers a huge advantage for fabrication and compaction. Table 1 shows that the microstructure could realize a high diffraction efficiency over 47.5% of the ±1$^{th}$ orders for RGB wavebands under TE polarization, and can meet the previous requirements of NEDs.

**CONCLUSIONS**

In this paper, we have proposed a design method and optical systems for an achromatic surface microstructure, which is composed with one triple-carved-sub-grating. Under TE polarization, the simulation results proved that the designed surface microstructure can diffract RGB lights achromatically. This surface microstructure could be duplicated and used in NEDs as the coupling optics, providing a lightweight and compact solution for the dispersion problem in diffractive waveguide technology, such as head-mounted display and wearable display in the future. This method and system could be one possibility to eliminate dispersion both horizontally and vertically simultaneously and independently. Because its light weight, compact size, easy duplication, cheap price, portability, and flexibility, it is a promising technology to bring about the Augmented Reality (AR) glasses similar to our normal glasses in our future daily life.

## REFERENCES

1. Rolland J, and Thompson K. See-through head worn displays for mobile augmented reality. *Proceedings of the China national computer conference* 2011; **7**: 28-37.
2. Kress B, and Starner T. A review of head-mounted displays (HMD) technologies and applications for consumer electronics. *SPIE Defense, Security, and Sensing* 2013; 87200A-87200A.
3. Martins R, Shaoulov V, Ha Y, and Rolland J. A mobile head-worn projection display. *Optics express* 2007; **15**: 14530-14538.
4. Takagi M, Miyao T, Totani T, Komatsu A, and Takeda T. Light guide plate and virtual image display apparatus having the same. U. S. Patent 2012.
5. Cheng D, Wang Y, Hua H, and Talha M M. Design of an optical see-through head-mounted display with a low f-number and large field of view using a freeform prism. *Applied optics* 2009; **48**: 2655-2668.
6. Andrew Maimone, Andreas Georgiou, Joel Kollin. Holographic near-eye displays for virtual and augmented reality. *ACM Trans Graph* 2017; **36**: 1-16.
7. Saarikko P. Diffractive exit-pupil expander for spherical light guide virtual displays designed for near-distance viewing. *Journal of Optics A: Pure & Applied Optics* 2009; 11: 065504.
8. Han J, Liu J, Yao X, and Wang Y. Portable waveguide display system with a large field of view by integrating freeform elements and volume holograms. *Optics express* 2015; **23**: 3534-3549.
9. Cheng D, Wang Y, Xu C, Song W, and Jin G. Design of an ultra-thin near-eye display with geometrical waveguide and freeform optics. *Optics express* 2014; **22**: 20705-20719.
10. Sarayeddine K, and Mirza K. Key challenges to affordable see-through wearable displays: the missing link for mobile AR mass deployment. *SPIE Defense, Security, and Sensing* 2013; **8720**: 87200D-87200D.
11. Aieta F, Kats MA, Genevet P, and Capasso F. Multiwavelength achromatic metasurfaces by dispersive phase compensation. *Science* 2015; **347**: 1342-1345.
12. Mohammadreza K, Francesco A, Pritpal K, Kats M A, Patrice G, David R *et al*. Achromatic metasurface lens at telecommunication wavelengths. *Nano Letters* 2015; 15: 5358.
13. Deng Z L, Zhang S, and Wang G P. A facile grating approach towards broadband, wide-angle and high-efficiency holographic metasurfaces. *Nanoscale* 2016; **8**: 1588-1594.
14. Mukawa H, Akutsu K, Matsumura I, Nakano S, Yoshida T *et al*. 8.4: distinguished paper: a full color eyewear display using holographic planar waveguides. *SID Symposium Digest of Technical Papers* 2008; **39**: 89-92.
15. Shi R, Liu J, Zhao H, Wu Z, Liu Y *et al*. Chromatic dispersion correction in planar waveguide using one-layer volume holograms based on three-step exposure. *Applied optics* 2012; **51**: 4703-4708.
16. Zhang N, Liu J, Han J, Li X, Yang F *et al*. Improved holographic waveguide display system. *Applied Optics* 2015; **54**: 3645-3649.
17. Simmonds M D, Valera M S. Display comprising an optical waveguide and switchable diffraction gratings and method of producing the same. 2017.
18. Crawford G P. Electrically switchable Bragg gratings. *Optics & Photonics News* 2003; 14: 54-59.
19. https://www.microsoft.com/microsoft-hololens/en-us.
20. Moharam M G, Gaylord TK, Grann E B, and Pommet D A. Formulation for stable and efficient implementation of the rigorous coupled-wave analysis of binary gratings. *Journal of the Optical Society of America A* 1995; **12**: 1068-1076.